\documentclass{article}

\usepackage[english]{babel}
\usepackage{natbib}

\usepackage[a4paper,top=3cm,bottom=3cm,left=2.5cm,right=2.5cm,marginparwidth=1.75cm]{geometry}

\usepackage{makecell}
\usepackage{amsmath}
\usepackage{amssymb}
\usepackage{adjustbox}
\usepackage{amsfonts}
\usepackage{graphicx}
\usepackage{authblk}
\usepackage{comment}
\usepackage{bbm}
\usepackage{booktabs}
\usepackage[colorlinks=true, allcolors=blue]{hyperref}
\usepackage{multirow}

\title{Communicating Complex Statistical Models to a Public Health Audience: Translating Science into Action with the FARSI Approach. 
}

\author{Mattia Stival, Lorenzo Schiavon, Gaia Bertarelli, Stefano Campostrini\\ 
{\small Department of Economics, Ca' Foscari University of Venice, San Giobbe, Cannaregio 873, 30121 Venice, Italy\\
\textbf{Corresponding authors}:\\ Mattia Stival, \texttt{mattia.stival@unive.it}; 
 Lorenzo Schiavon, \texttt{lorenzo.schiavon@unive.it}
}
}

\begin{document}

\maketitle

\section*{Abstract}
\textbf{Background}
 Effectively communicating complex statistical model outputs is a major challenge in public health. This study introduces the FARSI approach (Fast, Accessible, Reliable, Secure, Informative) as a framework to enhance the translation of intricate statistical findings into actionable insights for policymakers and stakeholders. 
 We apply this framework in a real-world case study on chronic disease monitoring in Italy.
\\ \noindent
\textbf{Methods}
 The FARSI framework outlines key principles for developing user-friendly tools that improve the translation of statistical results. 
 We applied these principles to create an open-access web application using {\sf R Shiny}, designed to communicate chronic disease prevalence estimates from a Bayesian spatio-temporal logistic model.
 The case study highlights the importance of an intuitive design for fast accessibility, validated data and expert feedback for reliability, aggregated data for security, and insights into prevalence population subgroups, which were previously unobservable, for informativeness.
\\ \noindent
\textbf{Results}
 The web application enables stakeholders to explore disease prevalence across populations and geographical area through dynamic visualizations. It facilitates public health monitoring by, for instance, identifying disparities at the local level and assessing risk factors such as smoking. Its user-friendly interface enhances accessibility, making statistical findings more actionable.\\ \noindent
\textbf{Conclusions}
 The FARSI framework provides a structured approach to improving the communication of complex research findings. By making statistical models more accessible and interpretable, it supports evidence-based decision-making in public health and increases the societal impact of research.


\section*{Keywords}
Bayesian Analysis, Chronic Disease Monitoring, Local Health Surveillance, Scientific Results Dissemination, Web application

\section{Background}

\subsection{Introduction}
In recent years, the application of advanced statistical models and machine learning in the social sciences has seen remarkable development \citep{gelman2011causality, grimmer2021machine}, driven by a variety of motivations. 
On one hand, there is greater availability of data and heterogeneous information sources \citep{benoit2016crowd,tian2021estimating}, as an important social and economic aspect to be considered in decision-making and political dynamics, with public and private institutions directly involved in data collection processes.
On the other hand, the availability of advanced computational infrastructures has boosted the development of methods for analyzing and managing large-scale data, such as, for instance,
topic analysis from large corpora of textual data \citep{dimaggio2013exploiting, roberts2020text}.
Thanks to complex models, it is now possible to monitor social phenomena in areas where, due to a lack of data or appropriate analytical and sampling strategies, there was previously a clear scarcity of information.
One other example is the development and application of Small Area Estimation (SAE) methods \citep{rao2015small, tzavidis2018start}, which have proven particularly useful in generating reliable estimates for small geographic areas or subpopulations, even when traditional survey sample data are insufficient. 

The aforementioned abundance of data and methods to analyze them have presented researchers across disciplines with a multitude of results and outputs that require careful consideration and interpretation. 
Although providing guidance on presenting statistical results to a general and non-technical audience has been a relevant topic in the literature for many years \citep{freeman1983presenting}, the abundance of new data and technologies has prompted critical reflection and introduced new challenges and opportunities to established practices in the presentation and interpretation of findings. Notably, this is exemplified by ongoing debates regarding the use of \emph{p-values} in scientific research \citep{Wasserstein02042016, ioannidis2019importance, matthews2021p}, as well as concerns about the \emph{reproducibility crisis} \citep{steneck2006fostering, tempini2018concealment}, with the latter being especially prominent in the social sciences \citep{baker2016reproducibility, trustinnumbers, fanelli2018science}.
In parallel, a burgeoning area of research in machine learning and artificial intelligence has focused on enhancing the explainability and communicability of complex ``black box" models \citep{minh2022explainable, dwivedi2023explainable}.
More broadly, challenges in science communication have gained attention across diverse scientific domains, as documented in various case studies and guidelines available in the literature \citep{fischhoff2013sciences, bucchi2014science,  national2017}. 
One example regards the case of \emph{vaccine hesitancy} in contrast to proven \emph{vaccine efficacy}, which  recognized as one of the top ten global health risks, with communication playing a pivotal role in overcoming it \citep{Feemster03052020,Bozzola31122020}. 
Similar concerns are discussed in the literature concerning COVID outbreak \citep{freeman2021communicating, woloshin2023communicating} or regarding climate changes issues \citep{york2021integrating, lewandowsky2021climate}.

As statisticians with different backgrounds and experiences, we asked ourselves how to communicate results of \emph{given}---and sometimes complex---statistical models to wider audiences, or to audiences which are different from those for which the statistical methodology was suited for.
Our effort is motivated by a real case study in healthcare, inspired by the recent work of \cite{stival2024bayesian} on public health monitoring in Italy,
leveraging repeated cross-sectional health surveys data.

While statistical methods are extremely powerful tools that are generally well understood by researchers in statistics and related fields, they could become cumbersome to interpret for people with few or none technical skills in the topic \citep{best2004more, testingBybetting, gigerenzer2007helping}, but that may have interest in the results of the scientific research. 
The non-communicability is indeed exacerbated by the fact that research outputs are not necessarily focused on  indicators that are well understood by the general public, due to specific research questions or the intrinsic complexity of the phenomena analyzed.
For instance, the case study presented here focuses on transforming the multiplicity of insights obtainable from the model by \cite{stival2024bayesian}, such as odds-ratios and correlation graphs, into a output which is more accessible for a broader audience 
interested in possible applications in term of public health. In fact, despite insightful,  interpreting these relationships is challenging due to the intricate dependencies between risk factors across populations. In addition, stake-holders involved in disease monitoring have generally limited access to the results provided, and this may be motivated by several reasons, including the impossibility in accessing journal articles from main publisher (e.g. due to high fees), restriction in accessing detailed sample compositions, which remain often confidential to mitigate privacy concerns, 
or even a limited knowledge on advanced statistical modeling techniques.

While we recognize the important role of publishing in peer-reviewed statistical journals, as a path to validate the work done from a methodological point of view, we interpret also the failure in translating results into easily accessible outcomes highlights as a general loss of information.
The goal of this paper is to suggest a possible way to mitigate this loss, in particular using the work by \cite{stival2024bayesian} as motivating example. 
We emphasize our focus on communicating the results of an already existing statistical model, which we consider as \emph{given}, meaning consolidated and validated.
Our interest does not lie in altering the analysis strategy to tailor the results for a different audience. 
Instead, we recognize the validity of the existing statistical model and the necessity of translating its output, with the aim of enhancing the impact of the research already conducted and its dissemination to a wider non-technical audience.
By doing so, we identify five key characteristics that dissemination tools should aim for. 
These characteristics, encapsulated by the acronym FARSI (Fast, Accessible, Reliable, Secure, Informative), represent principles that guided us the design and implementation of effective communication tools.
We will then refer to the term ``\emph{farsi}fication'' to describe the process of transforming statistical research outcomes into valuable insights tailored for a broader audience—distinct from the original scientific audience—while trying to align with the FARSI principles.
This process aims at meeting the needs of non-expert stakeholders, enhancing their accessibility and broadening the impact of scientific effort.
In the case study, 
we developed an open-access web-app using \texttt{Shiny} \citep{shiny}, a free and easy to use framework for building  web-app using \texttt{R} \citep{Rmanual}.

Similar frameworks have been proposed in other domains to guide the dissemination of information. For example, the FAIR principles (Findable, Accessible, Interoperable, and Reusable) \citep{}{wilkinson2016fair} have been widely adopted to enhance data sharing in scientific research, ensuring that datasets are structured for maximum usability. Likewise, the TIDieR checklist (Template for Intervention Description and Replication) \citep{hoffmann2014better} in healthcare research provides a standardized approach for reporting interventions to improve transparency and reproducibility. However, to the best of our knowledge, no comparable framework exists to specifically guide the communication of results from complex statistical models to the general population. By adopting the FARSI framework, we emphasize the importance of structured and user-oriented dissemination strategies for statistical modeling, bridging the gap between complex methodologies and accessible information for policies and decision-makers.


\subsection{Monitoring chronic diseases via repeated cross-sectional survey data}
Despite their great relevance in population monitoring, disease prevalence administrative data are often difficult to access \citep{ward2013estimating}. Sample data may represent an efficient alternative for inferring population prevalence and collecting further information possibly correlated. 
In this regard, various countries, including Italy, have developed notable methods for monitoring population health, which involve conducting frequent cross-sectional surveys at brief time intervals . These surveys, commonly known as BRFSS (Behavioral and Risk Factors Surveillance system), collect data on the well-being of individual participants, encompassing details about their behaviors, risk factors, and pertinent socio-demographic information (refer to  \cite{nelson2001reliability} and \cite{campostrini2005institutionalization} for discussions on data collection systems).
Given its considerable importance, various authors have analyzed the BRFSS datasets
\citep{AssafJRSSA15, ZHENG202374}. 
In Italy, PASSI (Progressi delle Aziende Sanitarie per la Salute in Italia, translated as ``Advancements of Local Health Units in Italy'') \citep{passi} is a surveillance system that, since 2008, collects sample data on the behavioral and risk factors of the Italian population in the age span of 18 to 69 years at the level of Local Health Units (LHU)  
\citep{baldissera2011peer}.
Recent works \citep{Pastore-passi22,andreella2023} have utilized the PASSI dataset to model the individual age-morbidity relationship and its association with socio-demographic risk factors.
However, the potential of the PASSI system extends beyond that, collecting valuable information at a local level that can be utilized to investigate the heterogeneity within the country. 
With sample data, operating at finer levels---i.e. by including cohort and location among structural exogenous information---could present challenges, as certain segments of the population may be underrepresented within the sample. Therefore, appropriate modeling is necessary to effectively leverage the available information across the entire dataset and to enhance the estimation process through the incorporation of prior knowledge.

A convoluted hierarchical Bayesian approach may offer a solution to address this challenge, enabling the integration of additional prior information to refine morbidity estimates.
In these concerns, \cite{stival2024bayesian}  propose a Bayesian multivariate hierarchical spatio-temporal logistic model for chronic disease diagnoses that can be applied not only at the national level but also at finer resolutions such as LHUs. In this paper we use this model as operating example in the application of the FARSI priciples. 

Formally, the model can be briefly expressed as follows.
Let $\boldsymbol{y}_i$ denote the $n_d$-variate binary response vector indicating if the disease $j \;(j=1,\ldots,n_d)$ has been diagnosed ($y_{ij} =1$) or not ($y_{ij} =0$) to the respondent $i\;(i=1,\ldots,n_s)$.
For each respondent $i$, an $n_p$-variate vector $\boldsymbol{x}_i$ of covariates, representing individual risk factors, is available. Covariates include information about sex, age, education level, economic status and smoking habits.
The authors proposed the multivariate logistic model
\begin{equation}
\label{eq:model}
    y_{ij} \sim \textnormal{Ber}(\pi_{ij}), \quad \pi_{ij} = \textnormal{logit}^{-1}(\eta_{ij}), \quad \boldsymbol{\eta}_i = B_i \boldsymbol{x}_i +\boldsymbol{\varepsilon}_i, 
\end{equation} 
where Ber$(\pi)$ denotes a Bernoulli random variable with mean $\pi$, $\boldsymbol{\eta}_i = (\eta_{i1},\ldots ,\eta_{in_d})$ store the linear predictors, and $\boldsymbol{\varepsilon}_i$ is the residual error term. 
Local levels are modeled through the specification of the coefficient matrix $B_i$ as dependent on the location and cohort to which the respondent $i$ belongs, such that both individual and environmental risk factors on the marginal probability of developing specific chronic diseases are considered. Additionally, the authors also includes a latent comorbidity indicator to model the residual error term $\boldsymbol{\varepsilon}_i$ to account for the correlation between diseases due to an individual's propensity for comorbidity. Further technical details of the model are presented in Section S1 of the Supplementary Material. 

\subsection{Insights and difficulties of a local-temporal morbidity model}
The wealth of information utilized in complex statistical approaches as that presented in  \cite{stival2024bayesian}  is counterbalanced by the inherent limitations faced by those who aim to fully leverage the data, such as policy makers in health,
and that is due to multiple concomitant causes.
On one hand, stakeholders involved in disease monitoring often have limited access to the results, either due to challenges in fully grasping the methodology or a lack of familiarity with statistical modeling techniques. On the other hand, researchers and end-users frequently encounter restrictions on accessing detailed sample compositions, which are crucial for validating results but must remain confidential to protect privacy. 

For example, a key feature of the model proposed by  \cite{stival2024bayesian}  is its ability to generate diverse insights by applying different analytical perspectives. 
The method enables analysis through various lenses, such as national, temporal, local, or individual viewpoints, though integrating these perspectives simultaneously can be challenging. 
\begin{figure}[ht]
    \centering
    \includegraphics[width=0.9\linewidth]{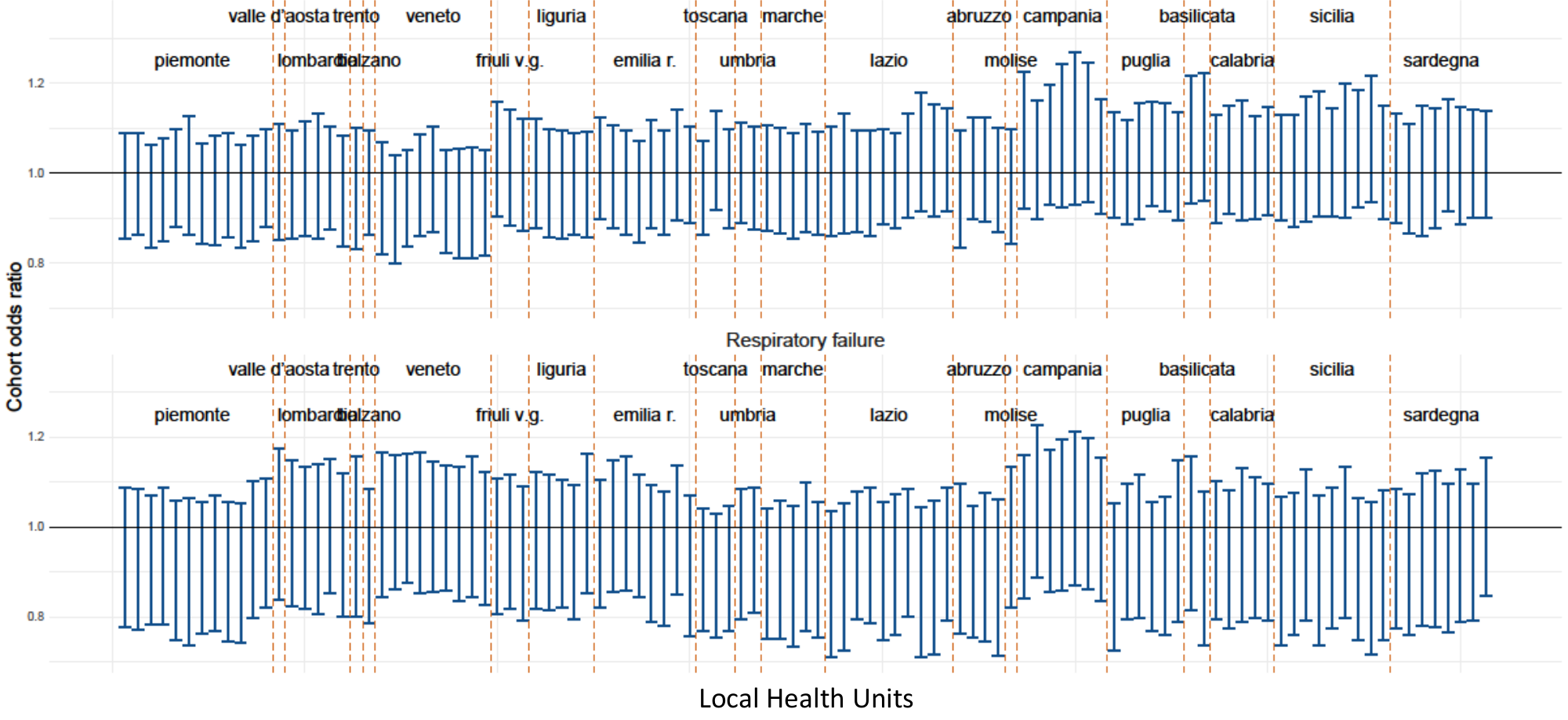}
    \caption{Example of posterior credible intervals of the predictive odds-ratio between two subjects born in two subsequent cohorts with identical remaining covariates.
    Vertical dashed lines delimit the Italian administrative regions.
    Horizontal line indicates no variations.
    The odds ratios are based on PASSI data considering cohorts 1956-1960 and defined as in \cite{stival2024bayesian}.
    }
    \label{fig:or-ex}
\end{figure}

Many insights from the paper, as is common in health statistical literature, are derived using odds ratios obtained through exponential transformations of the corresponding regression coefficients  $\exp\{\beta_{jh}\}$. 
To compare the heterogeneity of effects of specific risk factors across different local health units, the odds ratios are presented side by side, showing the values computed for different locations considered in the study.
Such results are generally difficult to represent and, consequently, to interpret. For instance, Figure \ref{fig:or-ex} presents an example of odds-ratio intervals for a risk factor of two diseases, computed for the 107 Italian LHUs.
 While this perspective may be useful for identifying patterns in the geographical variability of the specific impact of the risk factor on these diseases, it lacks a global view, as it focus on a specific aspects, overlooking several important aspects---such as, e.g., the residual interdependence between disease occurrences at the individual level, after controlling for risk factors. 
 Although this residual dependence, a crucial aspect in epidemiology and health statistics, may be captured by specific outputs of the model, 
 integrating these two findings---geographical variability and residual interdependence---is not straightforward, leaving the audience without a cohesive global message. 

A major focus in chronic disease monitoring often lies in analyzing prevalence in specific geographical areas at the population level rather than at the individual level. For example, health policymakers may be interested in studying phenomena such as morbidity compression---the reduction in the proportion of people within a certain age group suffering from chronic diseases in specific areas \citep{morbiditycompression, newton2021trends}---an important factor for resource allocation decisions \citep{Stallard01102016}.

To address this challenge, \cite{stival2024bayesian} analyzed the odds ratios between two respondents with identical characteristics but born in successive cohorts, similar to those reported in the example in Figure \ref{fig:or-ex}. These odds ratios highlight changes in the influence of risk factors across cohorts, which may be attributed to technological advancements or shifts in environmental risks for specific populations. While such analyses can help evaluate the underlying causes of cohort dynamics, they may be unnecessarily complex for describing aggregated phenomena, which are often the primary focus of routine public health monitoring.
Indeed, focusing solely on individual propensity for morbidity provides a limited perspective. Without access to detailed sample compositions, users struggle to fully understand and relate the observed cohort trends to the target population, since such trends have been derived for  a specific set of fixed covariates (e.g, a 62 years old wealthy highly educated female non-smoker). 
Even when a clear decreasing trend in individual propensity for chronic disease diagnosis over cohorts is observed \citep{Pastore-passi22, andreella2023, stival2024bayesian}---potentially due to technological advancements or improved treatment capabilities---an increase in the overall number of chronic disease cases within the population remains possible.
This apparent contradiction arises because improvements in health outcomes linked to risk factors may not align with population-wide trends. For instance, a reduction in the effects of risk factors on health outcomes could still coincide with worsening overall population health, or vice versa. These discrepancies can be attributed to demographic dynamics, such as increases in life expectancy, or the persistence of individuals failing to adopt protective behaviors. Shifting the perspective from individual-level analyses to aggregated dynamics is therefore essential to capture these broader trends and provide a \emph{fast} and \emph{informative} description of what is happening  ``overall''.

\section{Methods}

\subsection{\emph{Farsi}fication: translating results into valuable insights}

In designing tools for results dissemination, we aim to ensure that they meet several key characteristics, summarized by the acronym FARSI (Fast, Accessible, Reliable, Secure, and Informative). The framework represented by this acronym, captures multiple interconnected dimensions essential for effectively communicating the results of statistical models.
Operationally, it is a framework for identifying important features in the construction of dissemination tools, as well as for identifying their strengths and limits.
By adhering to these principles, we ensure that insights are conveyed in a way that is both comprehensible and actionable for diverse stakeholders, including policymakers, healthcare professionals, researchers, and the general public. 

In brief, the proposed acronym wants to refer to the following, from our point of view relevant, concepts. 
\begin{itemize}
\item \emph{Fast}:
the term  
 refers to all infrastructural aspects that guarantee the efficiency of the use of a tool, ensuring quick processing, intuitive access to results, and minimal computational delays.
\item \emph{Accessible}:
the term refers to a multifaceted requirement ensuring that the tool and its related information can be effectively used and understood by all potential users.

\item \emph{Reliable}:  the terms  refers to the consistency, accuracy, and trustworthiness of data, methods, and results, ensuring they are validated, reproducible, and derived from credible sources.

\item \emph{Secure}: the term  refers to the protection of data, privacy, and integrity by preventing unauthorized access, ensuring compliance, and mitigating misuse.

\item \emph{Informative}: the term
 refers to the ability of a tool to provide clear, relevant, and comprehensive insights that enhance users' understanding and support informed decision-making.

\end{itemize}
By adopting the FARSI framework, we emphasize the importance of structured and user-oriented dissemination strategies for statistical modeling. This approach wants to bridge the gap between complex methodologies and practical applications, ensuring that statistical insights translate into meaningful actions.
For policymakers, this means having access to clear, timely, and trustworthy information to design evidence-based interventions. For health professionals, it provides actionable guidance to improve patient outcomes and public health strategies. For researchers and analysts, FARSI-aligned tools foster collaboration, reproducibility, and transparency in scientific communication. Ultimately, this structured approach enhances informed decision-making, drives policy innovation, and improves societal outcomes. 

The five characteristics of the FARSI framework are here outlined in a general and concise way. Each application, however, requires that they are defined according to the specific purposes and characteristics of the problem addressed. 
Moreover, while the FARSI principles can be used as a guide for both developing tools, we highlight here how these can be used also for their evaluation by users, and self-evaluation by developers. 
In the following section, we will describe the specific choice we made for developing the tool. 
In the Results Section,  we will then present the functionalities of the tool, while we will present its use, together with a self-assessment, in the Discussion Section. 
Analyzing the tool according to the five Farsi dimensions allows us to highlight existing limitations and potentials, so as to make clear priorities and needs for new developments and maintenance.

\subsection{\emph{Farsi}fication of local-temporal morbidity model's results}

We present the application of FARSI framework in the considered health care example.
In particular, we explain the 
dissemination and communication process adopted to convey the results of  \cite{stival2024bayesian} to public health experts regarding chronic disease monitoring.
Specifically, each \emph{FARSI} characteristics is detailed in relation to the use that is required by the case study. In this case, \emph{farsification} is achieved by implementing a web-app.
While this part refers to the technical and methodological choices made to obtain the app, its detailed description is presented in Section \ref{sec:results}, and its practical use is discussed in Section \ref{sec:discussion}.
For ease of reading, 
the mathematical derivation of the presented output are reported in Section S2 the Supplementary Material.
\paragraph{Fast}
For the monitoring tool we aim to develop, the term \emph{fast} refers to two principal aspects.
On one hand, it emphasizes the intuitive need for technical infrastructure that enables efficient and rapid usage, thereby avoiding excessive delays when uploading results, generating visualizations, or creating tables.
This can be achieved by precomputing the most resource-intensive calculations, allowing users to access them as needed, or by implementing specific optimizations.
For example, the results provided in \cite{stival2024bayesian} are based on Monte Carlo computations using a sample of size $B = 3000$ from the parameter posterior distribution, as it is typical in Bayesian inference. Given the large number of parameters, processing these results---particularly in uncertainty quantification, which involves quantile calculations requiring extensive sorting operations---can be computationally demanding. Then, to enhance usability and responsiveness, we suggest thinning the sample in such cases, reducing its size sufficiently while maintaining representativeness to ensure reliable quantification.

On the other hand, \emph{fast} refers to the broader concept of a ready-to-use presentation of results, where only a few simple steps are needed to obtain the desired outputs, avoiding the need for multiple comparisons, ad hoc calculations, or advanced computing skills.
To address this aspect, in the case study discussed in this paper, 
one may translate the complex model outputs of a multivariate logistic regression models---such as odds ratios of conditional risk factor effects 
or residual error loadings---into quantities that are easily interpretable by all potential users, such as population prevalence, a widely used metric in health monitoring \cite[see][and references therein for recent examples in epidemiology and health policy evaluation]{yuan2021, peruga2021, steiner2023, sweeting2024}. 
Furthermore, these quantities should be pre-computed for all subpopulations of potential interest, allowing users to easily select and compare them through intuitive tools such as simple visualization filters, dropdown menus, or single-click options to add items. This approach ensures the process remains straightforward, accessible, and user-friendly for a broader audience.
This perspective is strongly related to the second adjective of the FARSI acronym, i.e. \emph{accessible}.

\paragraph{Accessible}

In the proposed solution, a key dimension of accessibility concerns the ability of the target audience to understand the presented results. In other words, while the underlying statistical model may be highly complex, the results must be presented on a scale familiar to the intended users. To address this, we recommend adhering to established best practices in data visualization \citep{telea2014data}.
In our case study, we provide plots with a limited number of clearly highlighted curves, differentiated by color.  Additionally, users are able to construct specific graphs and take advantage of features such as interactive zooming to explore the results in greater detail. Furthermore, tutorials and examples, available in accessible formats such as videos and PDFs, are included as readily available materials to guide users in interpreting and utilizing the outputs effectively.

In addition, to achieve accessibility, it is recommended to promote the scientific output through appropriate channels, make it available on open platforms, and accompany it with informative materials such as papers, technical reports, or tutorials. These materials should be released in open-access formats or stored in recognized open archives (e.g., Zenodo \cite{https://doi.org/10.25495/7gxk-rd71} and arXiv).
In the case study, we enhance accessibility by providing the outputs through a user-friendly web app, including a tutorial page, hosted on a freely accessible website: \url{https://sosta-apps.shinyapps.io/prevalence-curves/}. An early version of this web app was presented at the conference of the Italian Statistical Society, held in June 2024 in Bari, Italy, and was accompanied by an open-access short paper \citep{stival2024sis-app}. Additionally, a poster showcasing the web app was displayed at the public health conference PASSI e PASSI d'Argento: strumenti ad alta risoluzione per l'azione in Sanità Pubblica (\url{https://www.epicentro.iss.it/passi/incontri/convegno-10-dic-2024}).

\paragraph{Reliable} 

In the application, reliability should be firstly ensured by considering data coming from reliable sources and carrying out good validation practices.
Full referencing to data providers should be done, preferring official data providers and dataset with documentation, from those whose sources are unclear or poorly documented.

Second, it is essential that the tools used for analysis are validated. Validation should occur at different stages of the analysis and must be designed to demonstrate the instruments' ability to generalize results.
We distinguish between \emph{internal} and \emph{external} validation to emphasize the importance of using reliable tools at various stages of the analysis.
Internal validation refers to the ability of the model to generalize results using \emph{unseen} data that shares the same structure and properties as the sample used for estimation. This capability is typically assessed through techniques such as cross-validation or leave-one-out validation \cite[see, e.g.][]{vehtari2017practical}, where preferred models generally exhibit lower prediction errors.
In this case study, the model under consideration has been internally validated by the authors \citep{stival2024bayesian}, who compared its specification with alternative models using LOO-IC (Pareto-smoothed importance sampling Leave-One-Out Information Criterion) \citep{vehtari2017practical} and WAIC (Widely Applicable Information Criterion) \citep{watanabe2010}, demonstrating satisfactory performance.

On the other hand, external validation refers to the ability to generalize the results to \emph{unseen} data that differs in structure and properties from the sample used for estimation. 
In the context of population disease monitoring, as discussed in this paper, clinicians and epidemiologists often have access to administrative registers and disease-specific samples. These may include databases recognized by the scientific community as \emph{golden standards}, such as cancer registries \citep{mery2020population}. 
However, these resources may not always be fully accessible for the target population of interest, and they typically contain data aggregated at certain levels of analysis, requiring results aggregation at the population level.
Recent work by \cite{kennedy2023scoring} has proposed evaluation metrics for aggregate inferences in Bayesian predictions, accounting for the need to aggregate individual predictions to higher levels. 

As part of the external validation framework, we also incorporate validation by thematic experts who, with direct access to additional data (typically local) or specialized knowledge about the phenomenon being analyzed, can evaluate predictions from both qualitative and quantitative perspectives. The outputs presented in this paper were shown to the PASSI technical committee during the internal workshop ``Quattro PASSI per Ca' Foscari" (\url{https://www.unive.it/data/16437/1/88227}). In this context, the committee was asked to assess some estimated prevalence for areas or subpopulations within their areas of expertise as well as to judge its usefulness for PASSI surveillance system stakeholders.
Validity guarantees should ideally be supported by results in the literature, such as scientific articles recognized as credible by the intended audience. This is further strengthened by adhering to principles of transparency in data and methods, which facilitate additional validation. While full transparency is not always feasible due to data constraints, it is crucial to provide clear definitions of the tool and simulated examples that illustrate its strengths and limitations. However, achieving complete transparency---requiring access to all stages of the data process---may sometimes conflict with the imperative to protect citizens’ privacy.

\paragraph{Secure} 

Results must comply with privacy regulations and guidelines governing data collection and use. Online material accessible to external users should be crafted with safeguards in place to protect individuals' privacy by limiting access and storage to only necessary, and preferably aggregated, information.
In the context of population health monitoring, presenting results in aggregated rate formats---such as rates of prevalence for subgroups of the population---not only ensures privacy by avoiding the inclusion of individual-level details but also avoids disclosing the size of small populations, which could inadvertently reveal sensitive information about citizens.
When individual data is required, proper masking techniques must be employed to ensure anonymity.

\emph{Security} also pertains to preventing the misuse of the tool by users, that could intentionally (or not)  derive wrong or misleading results and interpretations. As developers, our ability to ensure proper use is inherently limited, as tools can be misinterpreted, distorted, or misused. \cite{blastland2020five} emphasize that in research areas such as climate change and COVID-19, people who are ``pre-emptively warned against attempts to sow doubt (known as prebunking) resist being swayed by misinformation or disinformation" (cfr. \emph{Nature} 587, page 364). As a preventive measure, it is advisable to develop clear guidelines, comprehensive documentation, and instructional videos to promote the correct use of the tool. Accurate referencing to relevant scientific materials can further aid users in understanding the tool and its intended applications.
Reliable results that are subject to misinterpretation risk rendering the tool ineffective or misleading, undermining its ultimate goal: \emph{informativeness}.

\paragraph{Informative} 

The tool should offer information that users either lack entirely, possess only partially, or would find prohibitively costly to obtain independently.
Moreover, the information provided should be as comprehensive as possible, rather than restricted to specific subgroups or areas.

In the proposed case study, we propose aggregating the outputs by population subgroups defined by all potential combinations of the risk factors considered by  \cite{stival2024bayesian}.
This approach provides insights into small population subgroups that are typically not included in registers, are too small for standard statistical analyses, or are not observed in sample data. It thus offers health monitoring stakeholders valuable information that was previously unavailable to them.
Estimating metrics for small population subgroups that are poorly or not at all represented in the sampled data poses challenges which are common in the SAE literature \citep{rao2015small, tzavidis2018start}. 

Formally, let $\boldsymbol{s}$ represent a set of conditions under which we aim to compute the prevalence for a new respondent $k$ in the population. 
We can split the covariates into two subsets $\boldsymbol{x}_k = (\boldsymbol{x}_{k\boldsymbol{s}}, \boldsymbol{x}_{k\bar{\boldsymbol{s}}})$, where $\boldsymbol{x}_{k\boldsymbol{s}} $ includes the covariates affected by the conditioning requirements (i.e., the risk factors of interest), and $\boldsymbol{x}_{\bar{\boldsymbol{s}}}$ includes those that remain unaffected by the specified conditions.
Informativeness is achieved by computing 
\begin{align*}
    p(\boldsymbol{y}^\star_k \vert \boldsymbol{x}_{k\boldsymbol{s}}, \mathcal{I}) = \sum_{\boldsymbol{x}_{k\bar{\boldsymbol{s}}} \in \mathcal{X}_{\bar{\boldsymbol{s}}}}  p(\boldsymbol{y}^\star_k \vert \boldsymbol{x}_{k\boldsymbol{s}},\boldsymbol{x}_{k\bar{\boldsymbol{s}}}, \mathcal{I}) p(\boldsymbol{x}_{k\bar{\boldsymbol{s}}} \vert \boldsymbol{x}_{k\boldsymbol{s}}, \mathcal{I}),
\end{align*}
where the summation (integration) is over $\mathcal{X}_{\bar{\boldsymbol{s}}}$, representing all possible finite combinations of the covariates unaffected by the conditions $\boldsymbol{s}$.
The quantity $p(\boldsymbol{y}^\star_k \vert \boldsymbol{x}_{k\boldsymbol{s}},\boldsymbol{x}_{k\bar{\boldsymbol{s}}}, \mathcal{I})$  represents the posterior predictive probability
that is derived using Monte-Carlo simulation as in the model by \cite{stival2024bayesian}. The term $\omega_{k\bar{\boldsymbol{s}}} =p(\boldsymbol{x}_{k\bar{\boldsymbol{s}}} \vert\boldsymbol{x}_{k\boldsymbol{s}}, \mathcal{I})$ indicates the probability that the \emph{new} respondent $k$ possesses the characteristics $\boldsymbol{x}_{k\bar{\boldsymbol{s}}}$, given the observed data $\mathcal{I}$ and characteristics $\boldsymbol{x}_{k\boldsymbol{s}}$.
As detailed in Section S2 of the Supplementary Material, we derived the quantity $\omega_{k\bar{\boldsymbol{s}}}$ by applying BART models \citep{JSSv097i01} to model the $\boldsymbol{x}_{k\bar{\boldsymbol{s}}}$ through the covariates $\boldsymbol{x}_{k\boldsymbol{s}}$.

\section{Results}
\label{sec:results}
\subsection{Web-app illustration}
To achieve our goal of making the results accessible to a broader audience, specifically policymakers and stakeholders of public health monitoring, we calculated prevalence rates for all combinations of covariates. These rates were visualized using line plots, illustrating prevalence trends over years (prevalence vs year) and across ages.
Consistently with Stival et al. \cite{stival2024bayesian}, the years monitored are pre-COVID, corresponding to the decade 2010--2020. 
Similarly, the age range considered spans the 50--65 age group, which is significant for monitoring chronic diseases in Italy \cite{Pastore-passi22}.
This focus enhances the \emph{informativeness} of the tool, as it provides an indicator of health trends in a population transitioning toward retirement and old age in the current (2020--2030) and upcoming decade (2030--2040).

\begin{figure}[th]
    \centering
    \includegraphics[width=0.45\linewidth]{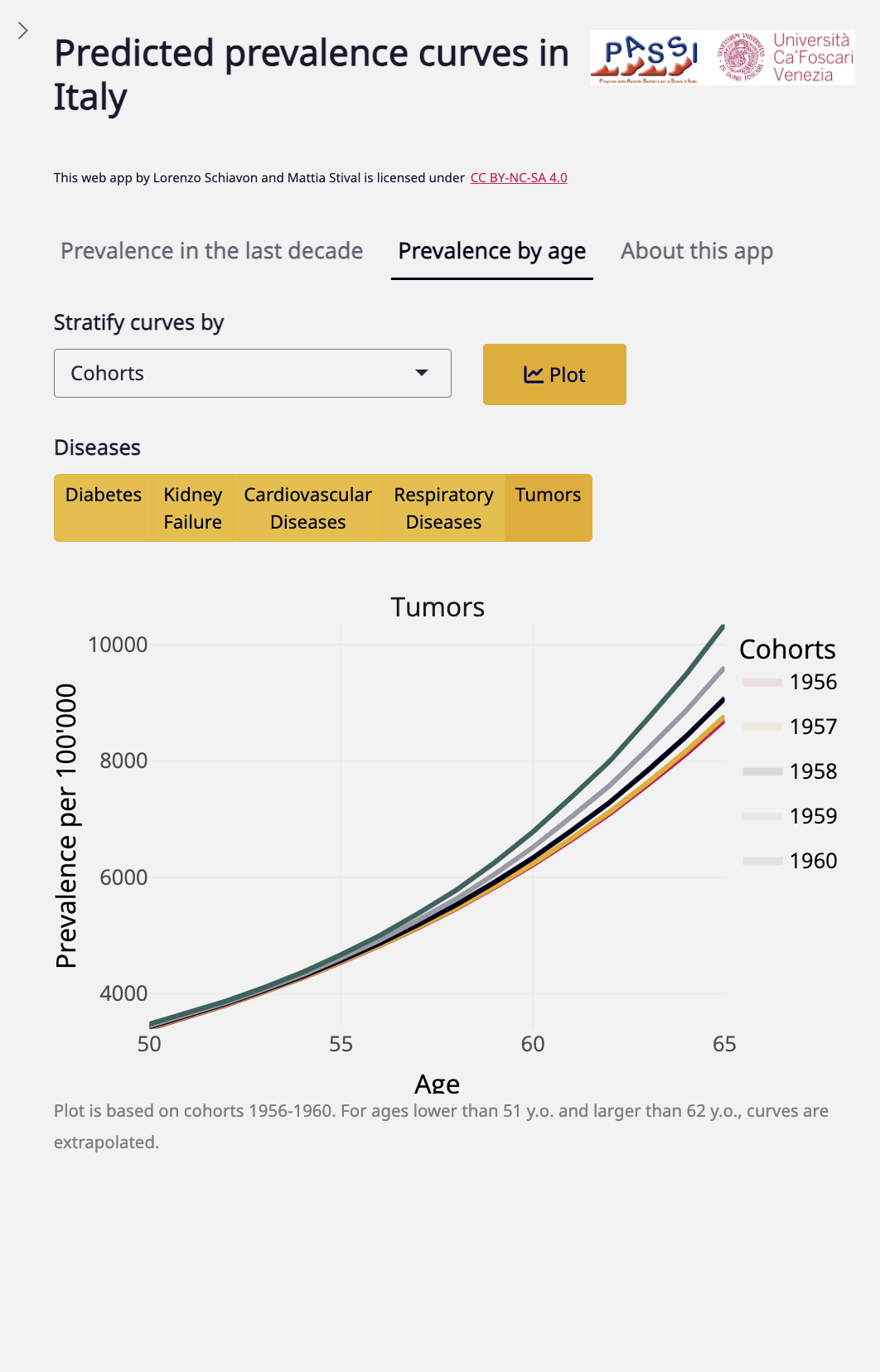}
    \caption{The mobile view of the web app showing age-prevalence for tumor diagnoses stratified by cohorts.}
    \label{fig:app-mobile}
\end{figure}

\emph{Accessibility} is ensured by making the plots available in a user-friendly web app hosted in the freely accessible website: https://sosta-apps.shinyapps.io/prevalence-curves/.
The web application infrastructure is developed using \textsf{Shiny} \citep{shiny}, an R package that simplifies the creation of interactive web applications directly within the R environment \citep{Rmanual}. 
Further accessibility requirements are also achieved through features like hovering over curves to see exact values, zooming in and out, and downloading graphs as PNG files.
The web app is also designed for mobile viewing, as shown in Figure \ref{fig:app-mobile}, ensuring fast and smooth access for users on the move.
In this section we present the web app structure, while specific examples of its use are reported in Section \ref{sec:discussion}, as well as the assessment of the app by using the FARSI framework.

\begin{figure}[t]
    \centering
\includegraphics[width=0.98\linewidth]{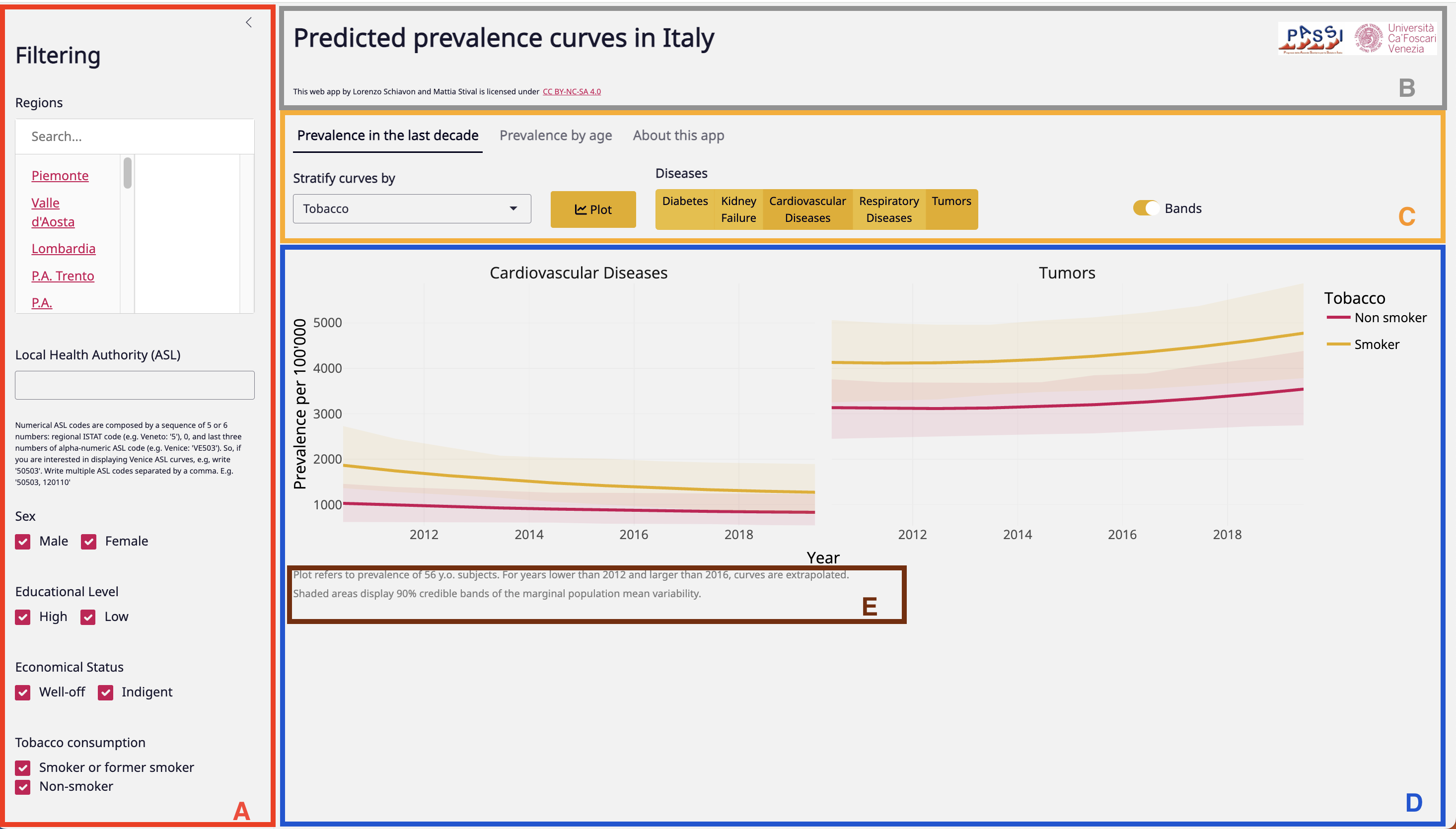}
    \caption{The structure of the web app after querying.}
    \label{fig:webapp}
\end{figure}

The structure of the website is illustrated in Figure \ref{fig:webapp}. The web page is divided into two macro sections. The first section, on the left (Section A), allows the user to select the population of interest, with options to filter by region or LHU of residence, sex, and other covariates used for estimation.
The second,  on the right, is the result display page (B--E), which can be expanded once the sub-population of interest is selected to ensure better visualization.
 
The display page is further divided into sections.
The header (Section B) includes the title of the app, along with reference information about the main parties involved in data collection and analysis: the PASSI surveillance system of the Italian Institute of Health (ISS), responsible for data collection and management, and the researchers at Ca' Foscari University of Venice, who carried out the analysis and developed the infrastructure.
This information helps the user recognize the key players involved in obtaining the results, thereby providing an implicit \emph{reliability} indicator associated with the institutions involved.
The license for the reuse of the results is also displayed there. We selected the ``CC BY-NC-SA 4.0" license, which allows users to ``Share" and ``Adapt" the material, provided they give appropriate credit, and under the conditions of ``Noncommercial" use and ``ShareAlike".

In Section C, the user can customize the visualization by selecting various options. The tab menu allows the user to choose the type of prevalence curve to display, either by years (as the example in Figure \ref{fig:webapp}) or by age. Figure \ref{fig:app-mobile} shows an example of the second tab.
Additionally, by clicking on a specific chronic disease, the user can select which disease to display. The option to stratify the curves by one of the covariates used in the analysis is also available, although the number of displayed curves is limited to a maximum of five.
We consider this a good balance between \emph{fast} usage and \emph{informativeness}, allowing users to compare risk factors while maintaining the ability to properly visualize and interpret the results.
The $90\%$ credible bands can be activated at the user's discretion. This aspect enhances the app's \emph{trustworthiness} \citep{trustinnumbers}, particularly for users familiar with statistical analysis, thereby contributing to the overall \emph{reliability} of the tool.
The tab ``About this app" provides informative material, including details about the authors, references to key papers used to obtain the displayed results, and tutorials. This page serves the dual purpose of enhancing both the \emph{accessibility} and informativeness of the app. For Italian users, there is an option to participate in an anonymous questionnaire aimed at collecting feedback on usage. Additionally, the questionnaire serves as a platform for reporting any inconsistencies in the displayed results or other issues encountered, allowing us to further verify and externally validate the findings.

Finally, the information banner (Section E) offers additional context to help users better understand the curves displayed in the main part of the website (Section D).

Operationally, \emph{fast} and \emph{secure} use of the app is ensured by precomputing the joint distribution
$
p(\boldsymbol{y}^\star_k, \boldsymbol{x}_k \vert \mathcal{D}) = p(\boldsymbol{y}^\star_k \vert \boldsymbol{x}_k, \mathcal{D}) p(\boldsymbol{x}_k \vert \mathcal{D}),
$
for any combination $h$ of the previously mentioned $107 \times 5 \times 12 \times 2^4 = 102,720$ possible covariate combination. Only the necessary information to generate the prevalence curves is stored on the server. For each combination of the covariates $ \boldsymbol{x}_k \in \mathcal{X}$, we store a sample of 300 particles drawn from 
$
p(\boldsymbol{y}^\star_k, \boldsymbol{x}_k \vert \mathcal{D}) = p(\boldsymbol{y}^\star_k \vert \boldsymbol{x}_k, \mathcal{D}) p(\boldsymbol{x}_k \vert \mathcal{D}),
$ which is used to online computing $p(\boldsymbol{y}^\star_k, \boldsymbol{x}_{k\boldsymbol{s}} \vert \mathcal{D})$
through simple summations when plots for the population group defined by 
 are requested by the user.
This approach ensures that no individual data are stored on the online server. Instead, only aggregate data in the form of random samples from the posterior distribution are stored, thus addressing privacy concerns effectively.

\section{Discussion}
\label{sec:discussion}
\subsection{Examples of use of the app  
for public health monitoring}

With the purpose of illustrating the use and usefulness of the app we present specific prevalence comparisons that may reveal valuable insights for Italian chronic disease monitoring.

\begin{figure}
    \centering
    \includegraphics[width=0.95\linewidth]{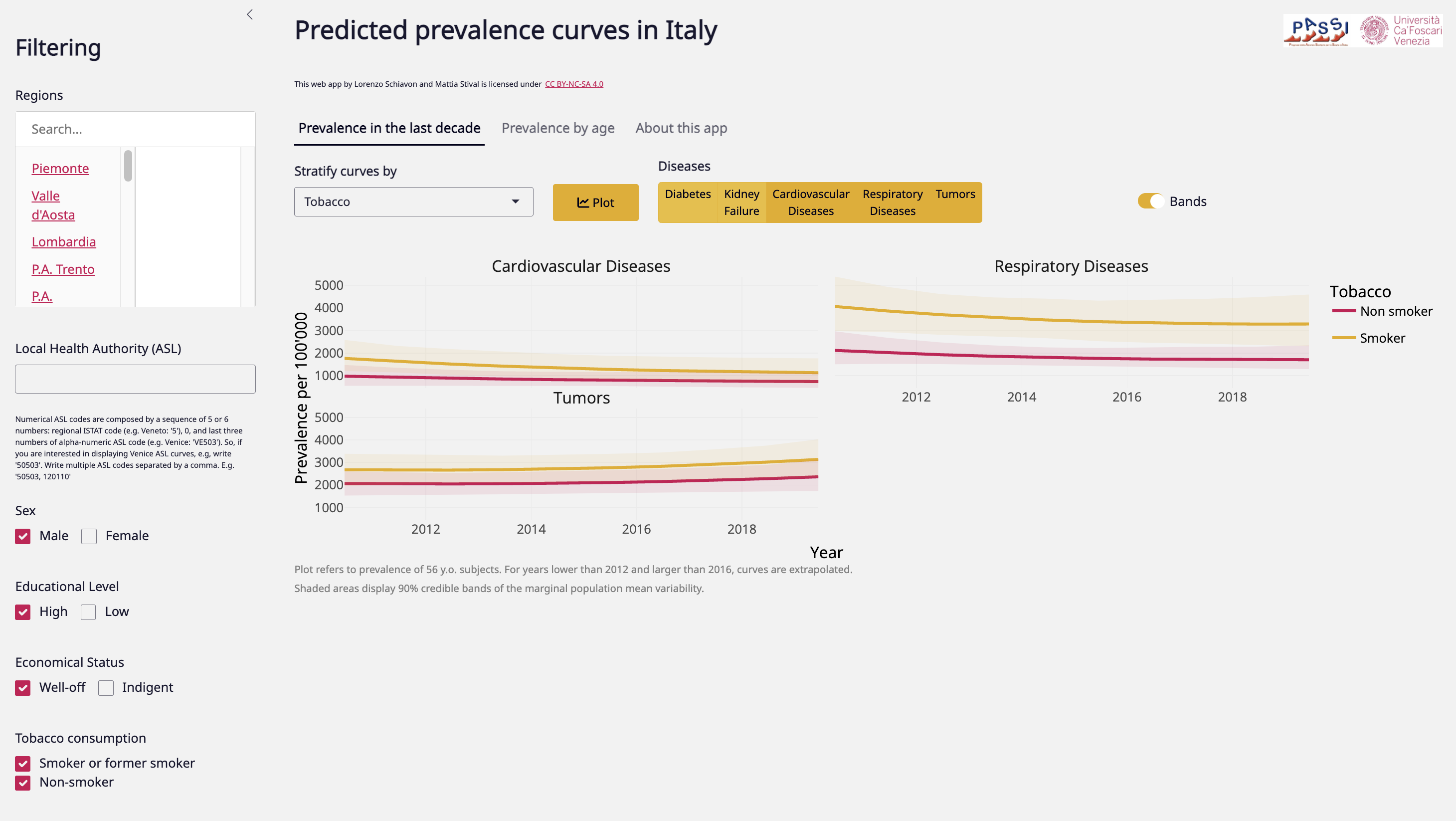}
    \caption{An example of a visualization  from the web app. 
    The plots display the prevalences of cardiovascular diseases, respiratory diseases and tumors, for smokers and non-smokers in the male, highly educated and well-off Italian population. Shaded areas represent the $90\%$ credible bands.}
    \label{fig:smoking}
\end{figure}

We first exploit the use of stratification with respect to possible risk factors on national population health, as the smoking behaviors.
Figure \ref{fig:smoking} displays the web app screen showing the prevalence curves over time for respiratory failures and (all) tumor diagnoses for smokers (and former smokers) and non-smokers with uncertainty quantification through $90\%$ credibility bands.
The figure also includes the filters used to select a population subgroup. Indeed, to avoid the inclusion of confounders, we focus on a tumor low-risk population (i.e., well-off, highly educated males). 
The negative effects of smoking on health are well-documented, with extensive literature regarding medical and epidemiological aspects \citep{giovino2002}, the development of preventive campaigns \citep{durkin2012}, health promotion initiatives, and studies on the economic impact of unhealthy behaviors \citep{possenti2024} and related policies on the healthcare system.
As expected, differences in estimated prevalence are substantial, ranging from approximately $500$ cases for cardiovascular diseases to $2,500$ cases for respiratory diseases per 100,000 individuals. 
In 2020, Istat estimated that approximately $22.3\%$ of the population aged 45--65 were ex-smokers, and $25\%$ were current smokers, within a total population of around $18$ million in this age group. In this roughly balanced population (smokers and ex-smokers vs. non-smokers), tobacco consumption is estimated to account for approximately $90,000$ additional cases of cardiovascular disease, $180,000$ cases of respiratory disease, and $500,000$ cases of tumors. 
Based on recent literature \citep{dalnegro2008, restelli2017, capri2017, mennini2024}, estimates of the average annual healthcare costs for patients with these chronic diseases range between €2,000 and €10,000. Consequently, the economic impact of tobacco consumption for the population aged 45--65 is projected to reach billions of euros annually.

By adjusting the filtering options and hovering over the curves with the mouse, one may notice that differences in prevalence magnitude remain similar even when conditioning on other covariate combinations or focusing on specific geographical areas. This observation suggests that the disparities are likely attributable to the effects of tobacco rather than spurious correlations with other risk factors.

While the previous example refers to the analysis of prevalence in the last decade, we now demonstrate the usefulness of the tab that reports prevalence curves by age  (see, e.g. \ref{fig:app-mobile}).
This perspective allows one to investigate, for instance, whether \emph{morbidity compression} occurs in Italy or in specific groups of regions.
Here, we use the term \emph{morbidity compression} to refer to the reduction over time in the number of people in the population suffering from a certain disease.

\begin{figure}
    \centering
    \includegraphics[width=0.95\linewidth]{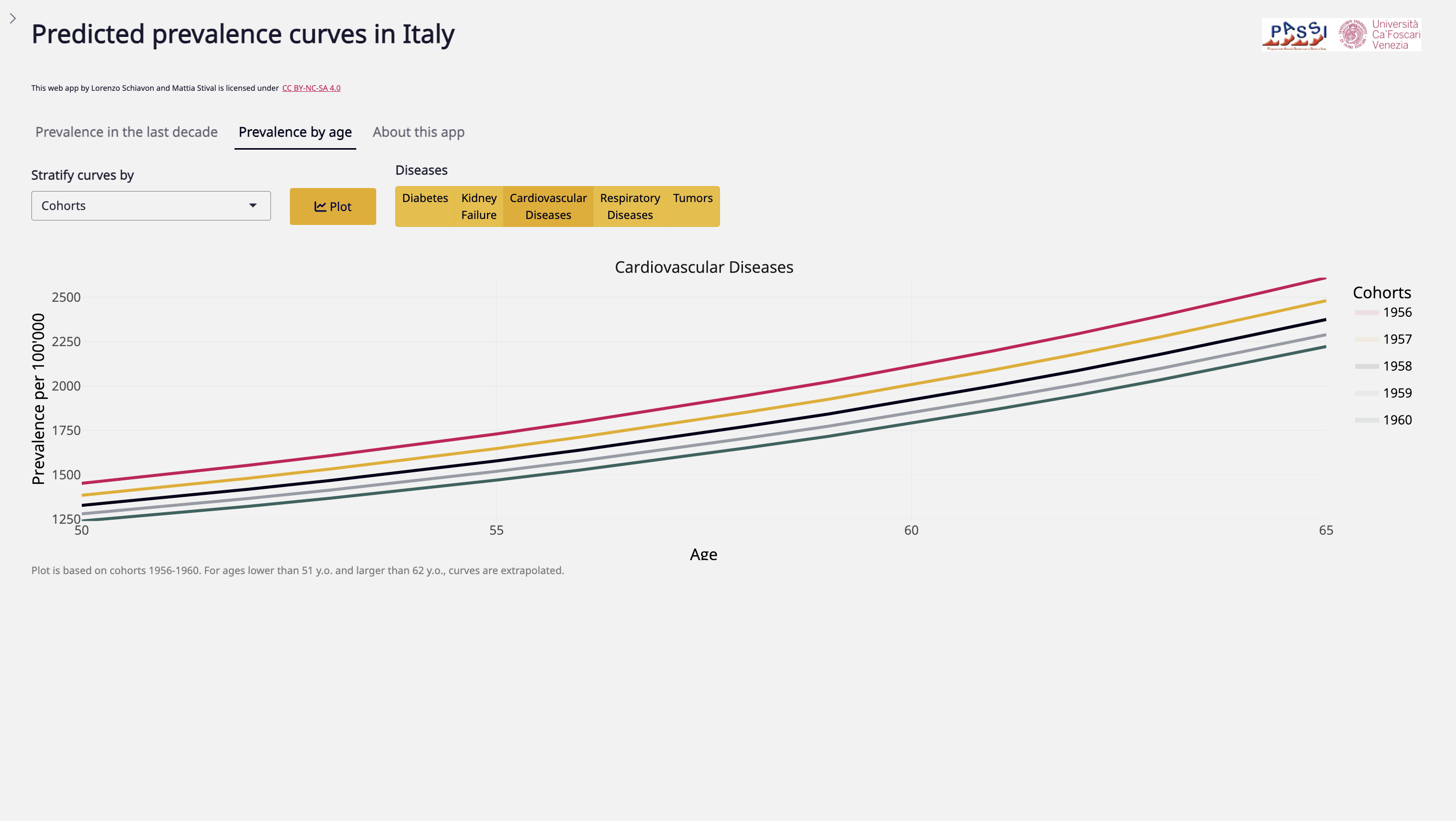}
    \caption{An example of a visualization from the web app. 
    The plot displays the age-prevalence curves of cardiovascular diseases stratified by cohorts. Higher curves correspond to older cohorts. }
    \label{fig:myocardic}
\end{figure}
\begin{figure}
    \centering
    \includegraphics[width=0.95\linewidth]{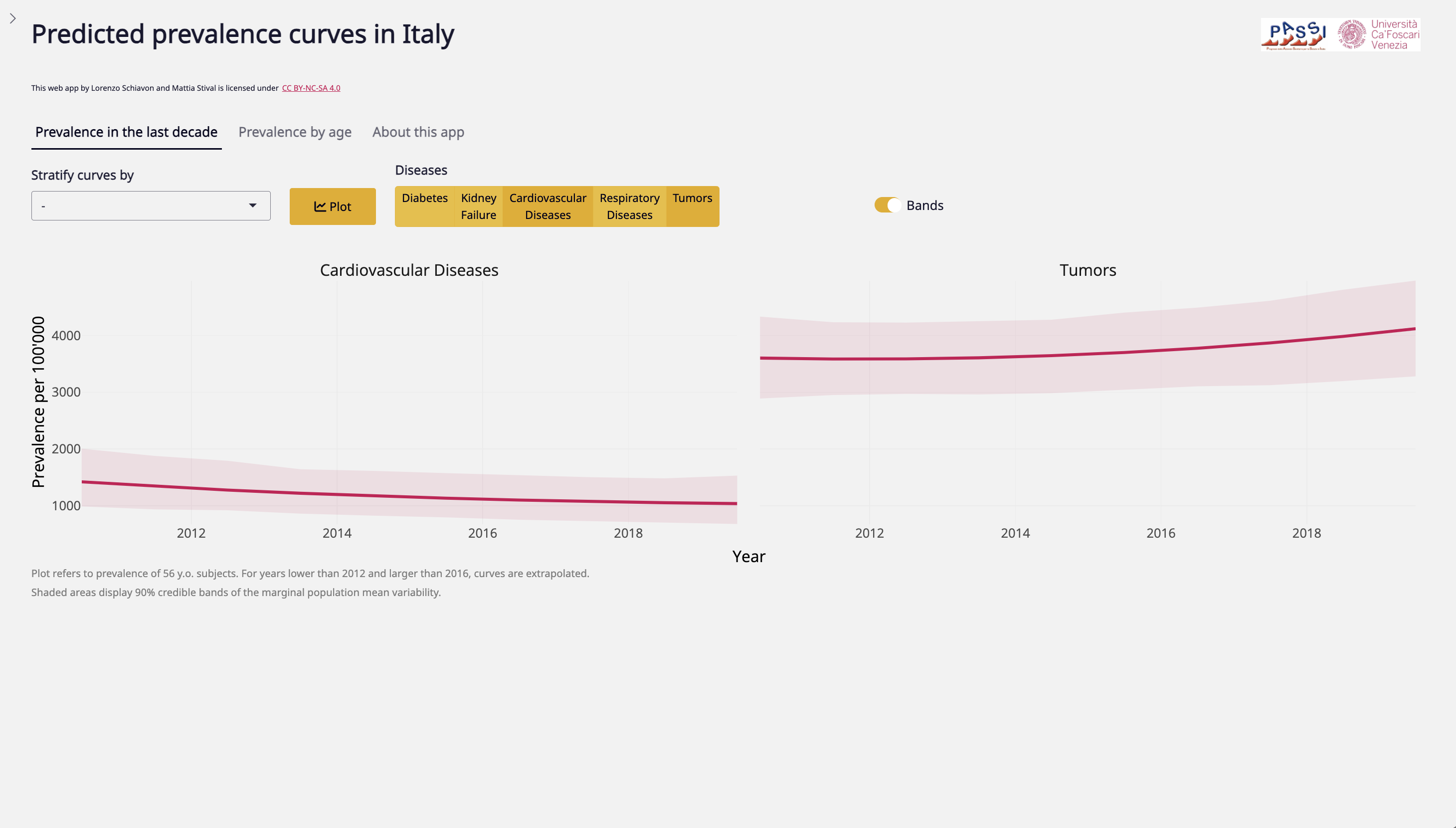}
    \caption{An example of a visualization from the web app. The plots report the prevalence curves over time for the Italian population illustrating the dynamics of morbidity compression. Shaded areas represent the $90\%$ credible bands.}
    \label{fig:morbidityCompressionDyn}
\end{figure}

\cite{Pastore-passi22} highlights a general morbidity compression in the prevalence of having at least one chronic disease.
Considering a more detailed view on specific diseases, 
\cite{stival2024bayesian} highlights disease-specific patterns. 
These patterns can also be visualized in the age-prevalence curves stratified by cohorts displayed by the app, as shown in Figures \ref{fig:app-mobile} and \ref{fig:myocardic}.
In particular, the figures show an increase in age morbidity curves across cohorts for tumors, whereas a decreasing trend is observed for cardiovascular diseases in the Italian population. 
Consistency between multiple views can be detected by switching tab and investigating the temporal dynamics of the prevalence over the years.
For instance, in  Figure \ref{fig:morbidityCompressionDyn}, we recognize
the opposite trend behaviours characterizing the two diseases. 
While the prevalence of cardiovascular diseases appears to decrease effectively, the number of diagnoses for tumors seems to increase over time.

At first glance, one might conclude that there is no morbidity compression for cancers, but rather an expansion. However, it should be noted that prevalence measures the proportion of people alive who have been diagnosed with a specific disease. An increase in prevalence does not necessarily imply an overall worsening of health. 
On the contrary, it could also reflect improved screening techniques and better treatments that lead to increased survival rates \citep{welch2010}.

\begin{figure}
    \centering
    \includegraphics[width=0.95\linewidth]{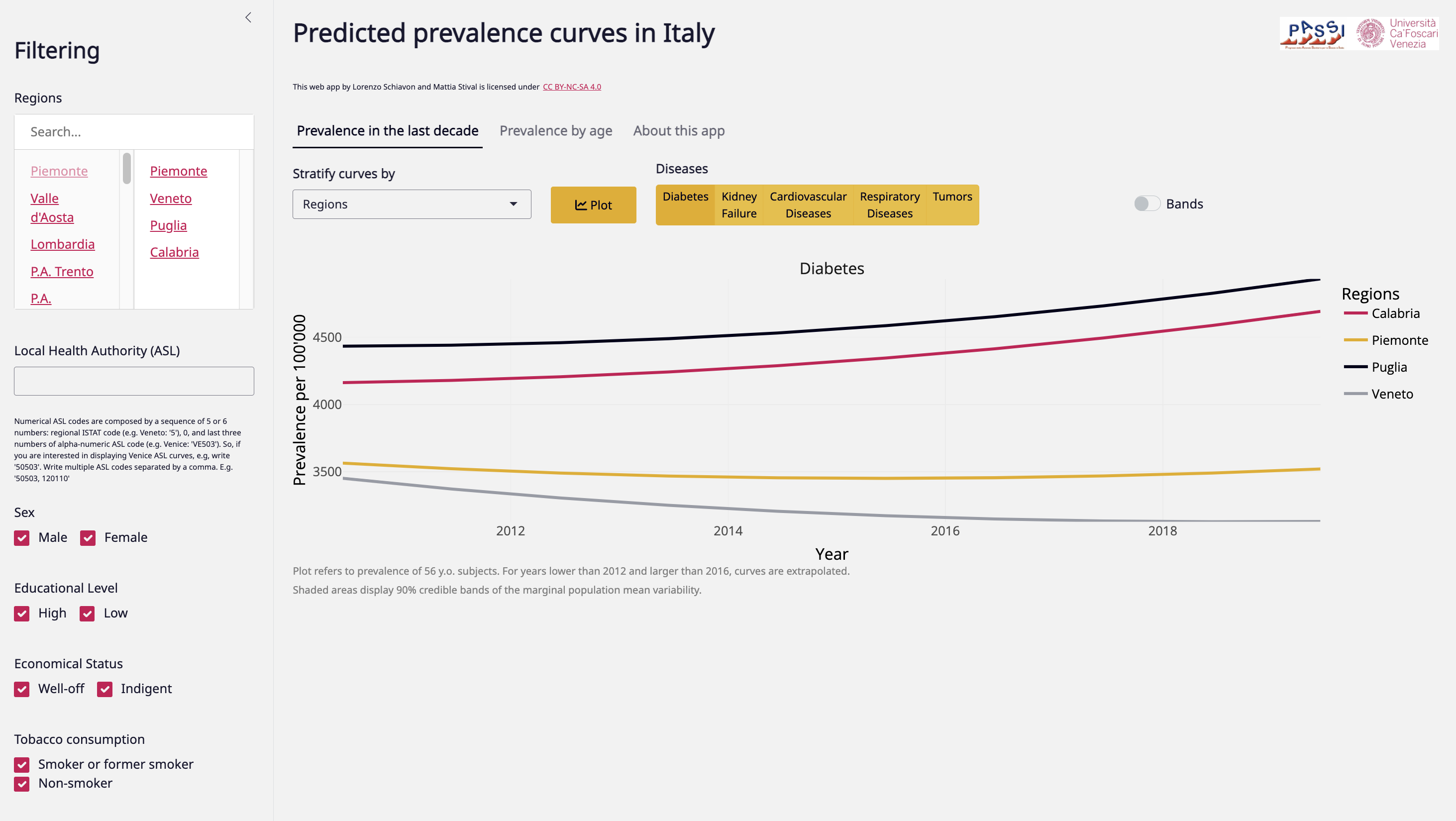}
    \caption{An example of a visualization from the web app showing temporal trends in diabetes prevalence for the regions of Piemonte, Veneto, Puglia, and Calabria. No filtering options are applied, so the curves reflect the estimated prevalence for the entire populations of these regions.}
    \label{fig:diabeteNordSud}
\end{figure}
Other than national comparisons, the app allows for comparisons at geographically restricted levels, such as regions or Local Health Units (LHUs). 
Figure \ref{fig:diabeteNordSud} displays, for example, the yearly trends of diabetes prevalence in the regions of Piemonte, Veneto, Puglia, and Calabria. In the figure, credible bands are omitted to ensure better visualization.
Northern and southern regions exhibit opposite trends, an aspect that appears robust even when alternative regions are selected. 
More specifically, while Puglia and Calabria show a slightly increasing trend in diabetes prevalence, Piemonte and Veneto demonstrate a decreasing trend. 
Notably, these differences in trends are not shared by other disease prevalences, with each disease displaying its own peculiarities. 
For instance, it is easy to verify through the app, by switching the displayed disease, that these regional trends seem to reverse when considering cancer prevalence. 
This suggests that Italy is not progressing uniformly in terms of health outcomes and that health inequalities are likely persisting or even widening.
By adjusting the filtering options, such as selecting more specific subpopulations to better control for potential confounders, it becomes possible to quantify how these dynamics may change under different conditions. 

\begin{figure}[th]
    \centering
    \includegraphics[width=0.95\linewidth]{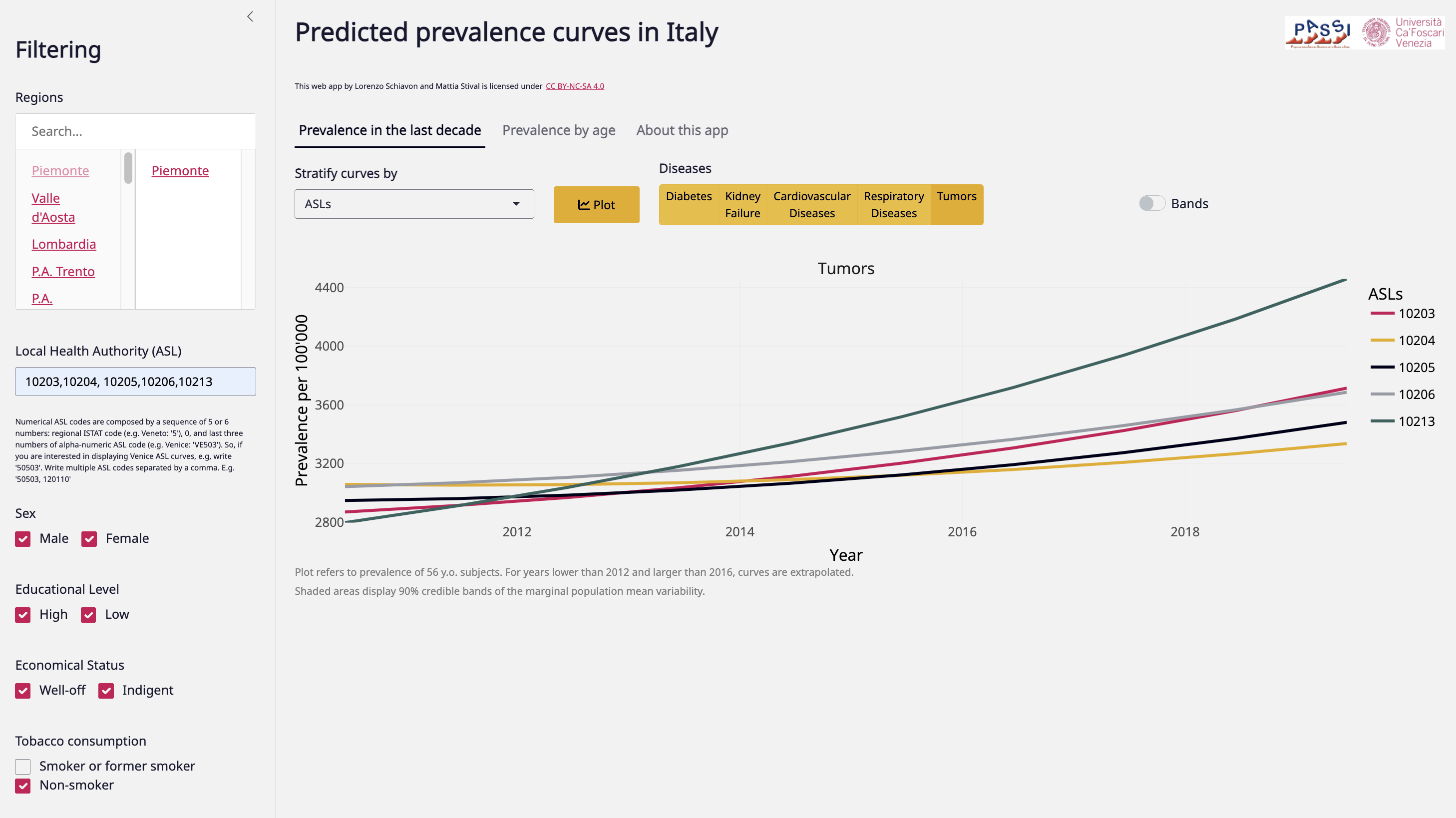}
    \caption{An example of a visualization from the web app. 
    Temporal trends in tumors prevalence for five LHUs in Piemonte. 
    To avoid the possible confounder due to tobacco consumption, analysis is conducted on non-smoker population, as it is shown by the filter toggle in the bottom left corner.}
    \label{fig:alessandriatumors}
\end{figure}

The app also enables the exploration of health dynamics within specific regions, offering insights for health professionals and policymakers interested in statistics at finer geographical levels, on which census data are less accessible.   
For instance, policymakers may be interested on studying diseases consequences in one of the 42 National Priority Contaminated Sites (NPCSs) \citep{isprambiente_contaminated_sites_2024} in Italy. 
In these cases, the app may help to compare Local Health Units (LHUs) within and beyond NPCSs to assess differences potentially linked to historical exposure to contaminants. 
Figure \ref{fig:alessandriatumors} illustrates the evolution of tumor prevalence for five LHUs in Piemonte, including Alessandria (10213), Torino province (10203, 10204, 10205), and Vercelli (10206). 
The prevalence curve of the Alessandria LHU shows a higher prevalence and a steeper trend compared to the others. 
While the differences may not be particularly strong in a strict statistical sense, they reinforce the importance of considering historical and environmental factors in health monitoring.
Indeed, Alessandria province has a notable history of asbestos pollution, particularly in Casale Monferrato (an asbestos NPCS) \citep{isprambiente_contaminated_sites_2024}, where Italy’s first and largest asbestos-cement factory operated \citep{comba2018mesothelioma}. 
Asbestos exposure is well-documented in the medical literature as a major risk factor for lung cancer, pleural mesothelioma and other types of cancer \citep{kwak2022environmental,clin2017cancer}
Consequently, Alessandria province has experienced a high incidence of mesothelioma due to occupational exposure \citep{ferrante2016}, with many cases of para-occupational exposure still being reported in the years of the analysis \citep{marinaccio2015}.

\subsection{Assessment of FARSI characteristics
}
In the FARSI framework for communicating results, \emph{informativeness} is the primary goal, supported by the interconnected characteristics of FARS. While an ideal tool would maximize all these aspects, practical applications and constraints often present trade-offs.

For instance, prioritizing speed may reduce reliability, as faster delivery might require less precise tools. In the web app, thinned MCMC samples were used to obtain credible intervals. Reducing the number of sample particles enables faster computation of quantiles and decreases the storage space required for outputs. However, this necessarily negatively impact the precision in quantile estimation, thereby decreasing the overall \emph{reliability} of the results.

Similarly, enhancing accessibility to improve usability may require simplifying complex phenomena, potentially reducing informativeness and reliability. For example, limiting simultaneous comparisons to a single stratification factor and a maximum of five curves per plot may improve readability but restrict analytical depth.

Finally, while this is not the case for our proposed tool, security concerns may sometimes limit the sharing of sensitive yet relevant findings.

Developing an effective tool requires balancing these factors through multidisciplinary collaboration. 
Technical teams ensure speed and security through robust infrastructure and data management, while statisticians and methodologists focus on reliability and informativeness.
Meanwhile, science communicators enhance accessibility and ensure secure information transmission.

Achieving \emph{farsification} demands a well-orchestrated effort tailored to the specific problem at hand. 
Evaluating success involves assessing whether initial objectives were met, by self-assessment, using user feedback, empirical testing, and iterative updates to refine the tool continuously.
For instance, Table \ref{tab:selfassestment2}  presents a self-assessment of the proposed tool, highlighting key areas for improvement. Each characteristic is assigned a self-evaluation score, along with a priority level for enhancements, making it easy to identify focus areas.
Additionally, for Italian users (the target audience), the app includes an evaluation survey with questions on usage and functionality in relation to FARSI characteristics.

\begin{table}
    \caption{\label{tab:selfassestment2} Table reporting the self-evaluation by the team that developed the app.}
    \centering 
     \renewcommand{\arraystretch}{1.2}
    \begin{tabular}{l p{3.8cm} p{3.8cm} l r}
    \hline
        \textbf{Characteristic}  &   \textbf{Actual Limitations} & \textbf{Needs to overcome limitations} & \textbf{Priority}  & \textbf{Self-Rating}\\
        \hline
         \emph{Fast}  
          & 1. Server memory constraints limit storage and computation. & 1. Upgrade to a proprietary server. & 1. Medium & 4/5 \\
          \hline
         \multirow{2}{*}{\emph{Accessible}}   & 1. App is only in English. & 1. Develop an Italian version. 
         & 1. Low & \multirow{2}{*}{4/5} \\
         & 2. Requires logical operations to filter data & 2. Provide tutorials on request. &        2. Low & \\
         \hline
         \emph{Reliable}   & 1. Few publicly available databases exist, with overly aggregated data.
          &   1. Collaborate with LHU operators for localized data. & 1. High & 3/5 \\
        \hline
         \emph{Secure}  & Not identified & None & None & 5/5   \\
         \hline
         \multirow{2}{*}{\emph{Informative}}   & 1. Data covers only the last decade and few cohorts. &  1. Update estimates with new data. & 1. Medium & \multirow{2}{*}{3.5/5} \\
        & 2. Stratification is limited to one variable. & 2. Explore alternative visualizations. &  2. Low 
         & \\
         \bottomrule
    \end{tabular}
    \label{tab:my_label}
\end{table}

\section{Conclusions}

In this paper, we build on previous research by integrating rigorous, complex, and sophisticated (for most of stakeholders) statistical modeling with effective communication strategies, ensuring that complex methods and programs in data collection and analysis maximize their societal impact through a structured and accessible dissemination process, particularly in the domain of public health. The FARSI framework provides a structured approach to making complex statistical outputs more comprehensible and actionable for a wider audience, and its implementation through our web application demonstrates the potential of such frameworks in bridging the gap between advanced statistical modeling and practical decision-making.

The web application, developed in accordance with the  FARSI principles, offers an innovative platform for exploring and visualizing trends in chronic disease prevalence across various demographic and geographical subgroups. By translating intricate model outputs into user-friendly insights, it enables policymakers, healthcare professionals, and researchers to make data-driven decisions more effectively. The positive reception of the app by stakeholders during the events where the app was presented highlights its potential to serve as a valuable tool in public health monitoring and intervention planning.

While this paper focuses on a specific case study, we believe the  FARSI framework could provide valuable insights for scientific communication in other domains. Future research could explore its applicability in different areas of public health, such as infectious disease surveillance, mental health monitoring, or environmental epidemiology. Additionally, adapting this approach to other fields beyond healthcare, such as social policy and economic modeling, may further demonstrate its versatility.

User engagement and feedback have played a crucial role in refining the application. The incorporation of feedback mechanisms, such as expert consultations, has allowed us to identify areas for improvement, including the need for additional data visualization features and more granular filtering options. Future updates will focus on enhancing these functionalities while maintaining the balance between ease of use and comprehensive data representation.

One key area for development is the scalability of the application. Integrating additional datasets, such as hospital records or regional health reports, could enhance the precision and applicability of the results. Furthermore, the potential for linking the platform with other health monitoring systems, including national and international databases, may provide a more holistic view of disease trends.

Despite its strengths, some challenges remain. One of the main limitations is the need for continuous updates to maintain data accuracy and relevance. Future work will explore automated data integration methods to streamline updates and ensure the tool remains a reliable resource for public health monitoring. Additionally, further research into Bayesian updating techniques and variational approximation approaches will be necessary to optimize real-time data incorporation.

The policy implications of this research are significant. By providing policymakers with accessible, data-driven insights into chronic disease trends, this tool can inform targeted health interventions, resource allocation, and preventive healthcare strategies. The ability to assess regional disparities and demographic risk factors allows for more effective public health planning and policymaking.

To effectively communicate statistical results, our approach breaks down the process into two stages: modeling and communication. In this way, even when using black-box models such as AI and machine learning, their outputs can be conveyed following the same communication principles if predictions are available. However, commonly used machine learning approaches like neural networks and gradient boosting do not inherently provide measures of uncertainty, making it difficult to trace back and interpret the sources of variability in the predictions. Addressing these limitations is crucial for ensuring \emph{reliability} in public health decision-making.

Ultimately, our approach highlights the importance of bridging the gap between statistical complexity and practical usability. By ensuring that research findings are not only accurate but also effectively communicated, we can enhance decision-making in public health and contribute to more informed policy development. The  FARSI framework serves as a model for future efforts in translating complex scientific research into actionable insights, fostering a more data-informed approach to societal challenges. Moreover, further collaborations with national health agencies and academic institutions could facilitate the expansion and refinement of this tool. Encouraging interdisciplinary approaches and knowledge-sharing will be key to ensuring its long-term success. By continuously evolving in response to user needs and technological advancements, the web application has the potential to become a standard reference for chronic disease monitoring in public health.

\section*{List of abbreviations}
\begin{itemize}
    \item BART: Bayesian Additive Regression Trees
    \item BRFSS: Behavioural and Risk Factors Surveillance System;
    \item FAIR: Findable, Accessible, Interoperable, and Reusable;
    \item FARSI: Fast, Accessible, Reliable, Secure, and Informative
    \item ISS: Istituto Superiore di Sanità (Italian National Health Institute)
    \item LHU: Local Health Unit;
    \item LOO-IC: Leave-One-Out Information Criterion;
    \item PASSI: Progressi delle Aziende Sanitarie per la Salute in Italia;
    \item PDF: Portable Document Format;
    \item SAE: Small Area Estimation;
    \item TIDieR: Template for Intervention Description and Replication;
    \item WAIC: Widely Applicable Information Criterion.
\end{itemize}

\section*{Declarations}

\subsection*{Ethics approval and consent to participate}
Not applicable.

\subsection*{Consent for publication}
Not applicable.

\subsection*{Availability of data and materials}
PASSI surveillance data can be accessed at \url{http://www.epicentro.iss.it/passi/}. 
The PASSI surveillance data that support the findings of this study  can be accessed at \url{http://www.epicentro.iss.it/passi/} and are available from the Italian National Institute of Health but restrictions apply to the availability of these data 
which were used under license for the current study, and so are not publicly available. Data are however available from the authors upon reasonable request and with permission of  National Institute of Health.

\subsection*{Competing interests}
The authors declare that they have no competing interests.

\subsection*{Funding}
LS and SC are  funded by the European Union - Next Generation EU, Mission 4 Component 2, within the NRRP (National Recovery and Resilience Plan) project ``Age-It - Ageing well in an ageing society" (PE0000015 - CUP H73C22000900006). 
    \\
	MS, GB and  SC   are funded by the European Commission grant 101136652. The five Horizon Europe projects, GO GREEN NEXT, MOSAIC, PLANET4HEALTH, SPRINGS and TULIP from the Planetary Health Cluster.
    \\
    The views and opinions expressed are only those of the authors and do not necessarily reflect those of the European Union or the European Commission. Neither the European Union nor the European Commission can be held responsible for them.
    
\subsection*{Authors' contributions}
\begin{itemize}
    \item MS: editing, writing, conceptualization, methodology development,  reviewing.
    \item LS: editing, writing, software \& app-development, 
    methodology development,  reviewing.
    \item GB: editing, writing, methodology development,  reviewing and supervision.
    \item SC: editing, reviewing and supervision. 
\end{itemize}

 \subsection*{Acknowledgements}
    The authors are grateful to the So.Sta. group (A. Andreella, A. Arletti,  M. Bertani, M. Marzulli, A. Pastore, M. Pittavino, S. Tonellato) at Ca’ Foscari University of Venice for their insightful comments.
    We also thank the Gruppo Tecnico PASSI (PASSI national coordination team at ISS) for their valuable feedback in the validation of the app, and the PASSI network interviewers for their diligent work in data collection.



\section*{Supplementary Material}

\section{The model}

Following \cite{stival2024bayesian}, for a generic respondent $i\;(i=1,\ldots,n_s)$, we denote with $\boldsymbol{y}_i$ the $n_d$-variate binary response vector indicating the presence or absence of $n_d$ possible diseases, with $y_{ij} =1$ if the disease $j$ has been diagnosed to the subject $i$. For each respondent $i$, an $n_p$-variate vector $\boldsymbol{x}_i$ of covariates, representing individual non-modifiable risk factors, is available, with $h$--th element $x_{ih}$  representing the value of covariate $h$.
Let Ber$(\pi)$ denote a Bernoulli random variable with mean $\pi$. Then, the authors propose the logistic model
\begin{equation}
\label{eq:model}
    y_{ij} \sim \textnormal{Ber}(\pi_{ij}), \quad \pi_{ij} = \textnormal{logit}^{-1}(\eta_{ij}), \quad \boldsymbol{\eta}_i = B_i \boldsymbol{x}_i +\boldsymbol{\varepsilon}_i.
\end{equation} 
In \cite{stival2024bayesian}, 
the residual error term of the model in Equation \eqref{eq:model} is specified as $\boldsymbol{\varepsilon}_i = \boldsymbol{\gamma} \epsilon_i$,  where the term $\epsilon_i$ is a scalar morbidity index accounting for propensity to co-morbidity of the respondent, scaled for each disease according to the parameter vector $\boldsymbol{\gamma}$. 

The $n_d \times n_p$ coefficient matrix $B_i$ is defined by the authors as a function $B_i=B(l_i, c_i)$ of the location $l_i$ and the cohort $c_i$ to which the respondent $i$ belongs to, with generic element $\beta_{jh}(l_i, c_i)$ representing the impact of the $h$-th covariate on the probability of disease $j$. 
Using a state space formulation, spatio-temporal dependence is included in the model by considering a dynamic-linear evolution for the terms in $B(l, c)$ and exploiting exogenous information available for the locations.
Trajectories of coefficients $\beta_{jh}(l_i, c_i)$ evolves linearly over cohorts $c_i$ according to the model
\begin{equation}
\label{eq:linear}
\beta_{jh}(l_i, c_i) = \beta_{jh}^0 + \lambda_{jh}^0 \xi_{jh}^0(l_i) +  (c_i - c_0) \lambda_{jh}^1 \xi_{jh}^1(l_i),
\end{equation}
where $c_0$ corresponds to the first cohort considered, and $\xi_{jh}^1(l_i)$ and $ \xi_{jh}^0(l_i)$ are location dependent random variables with standard Gaussian distributed marginals and correlation matrix $\mathcal{C}_{jh}^s$.
The correlation $\mathcal{C}_{jh}^s$ ($s \in \{0,1\}$) is expressed as a convex combination of different functions $\mathcal{K}_m(l, l^\prime)$ depending on external information on the locations. 
For instance, the authors considered an air pollution risk factor kernel $\mathcal{K}_4(l, l^\prime)=\exp\{-\mathcal{D}_4(l,l^\prime)\}$, with  $\mathcal{D}_4(l,l^\prime)$ a distance measure in terms of the average PM10 levels of location $l$ and $l^\prime$. By doing so, the authors assumed that similar locations in terms of pollution levels are characterized by higher correlation $\mathcal{C}_{jh}^s$ and, as a consequence, by higher similarities among regression coefficients.
This formulation allows for the inclusion of exogenous sources of variations to explain dependence between locations.
Refer to \cite{stival2024bayesian}  for further details on their proposed approach.

\section{Computing population diseases prevalence from individual probabilities}
 In alignment with the granularity of the available information, the model in  \cite{stival2024bayesian} focuses on capturing the relationship between risk factors and morbidity at the respondent level. Although this perspective is common in the literature within this field \citep{Pastore-passi22, AssafJRSSA15}, it limits the immediate interpretability of the results in terms of prevalence or broader population trends.


However, aggregation to higher levels is still possible by applying suitable transformations of the output quantities.
One of the advantage of Monte Carlo Bayesian inference is the possibility 
to easily obtain a sample from the posterior predictive probability
\begin{align*}
    p(\boldsymbol{y}^\star_k \vert \boldsymbol{x}_k, \mathcal{I}) = \int p(\boldsymbol{y}^\star_k \vert \boldsymbol{x}_k, l_k, c_k, \boldsymbol{\theta}) p(\boldsymbol{\theta} \vert \mathcal{I}) \text{d}\boldsymbol{\theta}.
\end{align*}
In this equation, $\boldsymbol{y}^\star_k$ represents a binary vector of observable outcomes for a \emph{new} respondent $k$ characterized by location $l_k$, cohort $c_k$, and the individual risk factors $\boldsymbol{x}_k$ considered in the model. These risk factors include demographic characteristics (sex and age) and other risk factors  (educational level, economic status, and smoking status).
The measure $p(\boldsymbol{\theta} \vert \mathcal{I})$ represents the posterior distribution of the parameters $ \boldsymbol{\theta}$ given the data $\mathcal{I}$, representing the information collected in the survey and used for estimation.
By marginalization, it is possible to compute $ p({y}_{kj}^\star =1\vert \boldsymbol{x}_k, l_k, c_k, \mathcal{I})$, which represents the posterior probability of receiving a diagnosis for disease $j$ for an individual with covariates $\boldsymbol{x}_k$, residing in location $l_k$, and born in cohort year $c_k$.
For instance, we may be interested in the posterior probability of being diagnosed with diabetes for a $55$-year-old, wealthy, highly educated, non-smoker male respondent, born in $1962$, and residing in the Venice LHU area.
In $2017$ (when a respondent born in $1962$ would be $55$ years old), this probability can be interpreted as an estimate of the prevalence of diabetes for individuals with the same exact characteristics.

Presenting model outputs in terms of prevalence offers significant advantages for communication. First, it shifts the focus from parameter estimation to model predictions, providing a single, aggregated output for a given group of people and a specific disease. 
This output synthesizes insights from multiple parameters, simplifying interpretation. It is indeed discussed in Bayesian literature---especially in prior elicitation literature,  \citep[see for example][]{mikkola2024prior,hartmann2020flexible} and references therein--how focusing in directly \emph{observable} aspects, rather than \emph{unobservable} parameters, offers an interpretive advantage.
This approach is especially valuable in highly parameterized models, where comparing and interpreting parameter estimates across numerous covariate combinations can be complex.
Second, prevalence is a widely recognized and commonly used metric in health monitoring. Its familiarity enables even a general, non-specialist audience to readily interpret and derive meaningful insights from the results.
To obtain the posterior probability of being diagnosed, it is necessary to specify all the characteristics of the respondent of interest.  
In the app, we consider $n_a = 107$ LHUs, $n_c = 5$ birth cohorts, an age span of $16$ years, and $4$ binary covariates, resulting in $107 \times 5 \times 16 \times 2^4 = 136,960$ possible covariate combinations, compared with approximately $22,000$ statistical units used for estimation.
This implies that probabilities can be obtained for very small subgroups of the population, including subgroups that are rare---or even nonexistent---in the sample data.
However, with such a high dimensionality, the resulting fragmentation of the sample makes it impractical to analyze population dynamics effectively, as it becomes difficult to grasp the relative weight or significance of each subgroup within the overall population.
On the other hand, potential users of the results, such as policymakers, may be more interested in less granular quantities. For example, they might seek information on the diabetes prevalence in $2017$ among $55$-year-old males residing in the Venice LHU area, without considering additional details like smoking status, educational attainment, or economic level.
Similarly, users may focus on broad metrics, such as cancer prevalence among smokers and non-smokers across a larger geographical region, such as an entire region (e.g., Veneto). 

To move from individual-level predictions to larger scales, it is necessary to adapt the results using averaging and post-stratification techniques \citep{gelmanPoststrat,parker2023comprehensive}.
More specifically, let $\boldsymbol{s}$ represent a set of conditions under which we aim to compute the prevalence. 
We can split the covariates into two subsets $\boldsymbol{x} = (\boldsymbol{x}_{\boldsymbol{s}}, \boldsymbol{x}_{\bar{\boldsymbol{s}}})$, where $\boldsymbol{x}_{\boldsymbol{s}} $ includes the covariates affected by the conditioning requirements (i.e., the risk factors of interest), and $\boldsymbol{x}_{\bar{\boldsymbol{s}}}$ includes those that remain unaffected by the specified conditions.
The interest is in
\begin{align*}
    p(\boldsymbol{y}^\star_k \vert \boldsymbol{x}_{k\boldsymbol{s}}, \mathcal{I}) = \sum_{\boldsymbol{x}_{k\bar{\boldsymbol{s}}} \in \mathcal{X}_{\bar{\boldsymbol{s}}}}  p(\boldsymbol{y}^\star_k \vert \boldsymbol{x}_{k\boldsymbol{s}},\boldsymbol{x}_{k\bar{\boldsymbol{s}}}, \mathcal{I}) p(\boldsymbol{x}_{k\bar{\boldsymbol{s}}} \vert \boldsymbol{x}_{k\boldsymbol{s}}, \mathcal{I}),
\end{align*}
where the summation (integration) is over $\mathcal{X}_{\bar{\boldsymbol{s}}}$, representing all possible finite combinations of the covariates unaffected by the conditions $\boldsymbol{s}$.
The term $\omega_{k\bar{\boldsymbol{s}}} =p(\boldsymbol{x}_{k\bar{\boldsymbol{s}}} \vert\boldsymbol{x}_{k\boldsymbol{s}}, \mathcal{I})$ indicates the probability that the \emph{new} respondent $h$ possesses the characteristics $\boldsymbol{x}_{k\bar{\boldsymbol{s}}}$, given the data and characteristics $\boldsymbol{x}_{k\boldsymbol{s}}$.
For instance, if the covariates in the subset $\boldsymbol{x}_{\boldsymbol{s}}$ correspond to the demographic characteristics used in the survey sampling design scheme, under ignorable design, the terms $\omega_{k\bar{\boldsymbol{s}}}$ simply reduce to a transformation of the sampling weights (e.g., the inverse transformation of the weights).

However, limiting $\boldsymbol{x}_{k\bar{\boldsymbol{s}}}$ to only the demographic characteristics used in the survey sampling design scheme may be overly simplistic, and scarcely informative for a potential user. This highlights the need for a more robust strategy to obtain $\omega_{k\bar{\boldsymbol{s}}}$. 
 \cite{stival2024sis-app} address this by estimating  these terms using the relative frequencies observed in the sample 
$\omega_{k\bar{\boldsymbol{s}}} = \{\sum_{i=1}^{n_s} \mathbbm{1}(\boldsymbol{x}_{i \bar{\boldsymbol{s}}}=
\boldsymbol{x}_{k\bar{\boldsymbol{s}}})\mathbbm{1}(\boldsymbol{x}_{i \boldsymbol{s}}=\boldsymbol{x}_{{k\boldsymbol{s}}})\}/\{\sum_{i=1}^{n_s} \mathbbm{1}(\boldsymbol{x}_{i {\boldsymbol{s}}} =\boldsymbol{x}_{k{\boldsymbol{s}}}) \}$, with $\mathbbm{1}(a=b)=1$ if $a=b$ and $0$ otherwise. 
While straightforward, this strategy has two significant shortcomings.
First, it restricts frequency computation to covariate combinations that are observed in the sample.
This results in assigning an unlikely value of $0$ to unobserved combinations, which arise due to sample limitations. Such an approach inherently limits inference to a narrow set of covariate combinations, neglecting the possibility of plausible but unobserved configurations.
Second, the approach lacks an adequate mechanism for uncertainty quantification, which is essential for ensuring the reliability of estimates reported to the general public \citep{blastland2020five}.

To obtain estimates of ${\omega}_{k\bar{\boldsymbol{s}}}$ with uncertainty quantification, we propose modeling $p(\boldsymbol{x}_k \vert \mathcal{I})$ within the Bayesian framework and by considering the decomposition 
\[
p(\boldsymbol{x}_k \vert \mathcal{I}) = p(\boldsymbol{x}_{k{\boldsymbol{d}}}, \boldsymbol{x}_{k\bar{\boldsymbol{d}}} \vert \mathcal{I}) = p(\boldsymbol{x}_{k{\boldsymbol{d}}} \vert \mathcal{I}) p(\boldsymbol{x}_{k\bar{\boldsymbol{d}}} \vert \boldsymbol{x}_{k{\boldsymbol{d}}}, \mathcal{I}),
\]
where $\boldsymbol{x}_{k{\boldsymbol{d}}}$ includes the demographic information (year of interview, age, sex, and LHU) of the $h$-th respondent, while $\boldsymbol{x}_{k\bar{\boldsymbol{d}}}$ includes the remaining individual risk factors (smoking status, educational level, and economic status).
By doing so, $p(\boldsymbol{x}_{k{\boldsymbol{d}}} \vert \mathcal{I})$ can be obtained as a deterministic transformation using official census data (based on ISTAT data).
A posterior sample from $p(\boldsymbol{x}_{k\bar{\boldsymbol{d}}} \vert \boldsymbol{x}_{k{\boldsymbol{d}}}, \mathcal{I})$ is then derived using Bayesian Additive Regression Tree (BART) models for categorical and discrete data \citep{JSSv097i01}.
Specifically, the response variable is defined by the eight categories resulting from all possible combinations of the three binary risk factors considered in \cite{stival2024bayesian} : smoking status (\emph{smoker or former smoker} vs. \emph{non-smoker}), economic level (\emph{low} if the respondent reports difficulty making ends meet, \emph{high} otherwise), and educational level (\emph{low} if the respondent has not obtained at least a high school diploma, \emph{high} otherwise).
The BART model is then used to obtain a posterior sample of $p(\boldsymbol{x}_{k\bar{\boldsymbol{d}}} \vert \boldsymbol{x}_{k{\boldsymbol{d}}}, \mathcal{I})$ using the demographic information $\boldsymbol{x}_{k{\boldsymbol{d}}}$ as covariates. Once a sample from $p(\boldsymbol{x}_k \vert \mathcal{I})$ is obtained, any posterior sample of ${\omega}_{k\boldsymbol{s}}$ can be derived through  marginalization. 
The use of BART is motivated by its flexibility in modeling complex relationships and without the need, in this case study, to provide an interpretation to them. While alternative approaches may be worth exploring, they are beyond the scope of this paper.

\bibliography{references}

\end{document}